\documentclass[letterpaper,10pt]{article}
\usepackage[utf8]{inputenc}
\usepackage[margin=0.75in]{geometry}

\makeatletter
\renewenvironment{abstract}{%
    \if@twocolumn
      \section*{\abstractname}%
    \else 
      \begin{center}%
        {\bfseries \abstractname\vspace{\z@}}
      \end{center}%
      \quotation
    \fi}
    {\if@twocolumn\else\endquotation\fi}
\makeatother

\usepackage{authblk}
\usepackage{braket}
\usepackage{amsmath}
\usepackage{amssymb}
\usepackage{mathrsfs}
\usepackage{physics}
\usepackage{graphicx}
\usepackage{float}
\usepackage{times}
\usepackage{multicol}
\usepackage{subcaption}
\usepackage{fancyhdr}
\usepackage{changepage}

\title{\textbf{Synchronized neutrino communications over intergalactic distances}}

\author[1]{A. D. Santos}
\author[1,2]{E. Fischbach\thanks{Corresponding author. ephraim@purdue.edu.}}
\author[2]{J. T. Gruenwald}
\affil[1]{\textit{Department of Physics \& Astronomy, Purdue University, West Lafayette, IN 47906, USA}}
\affil[2]{\textit{Snare, Inc., West Lafayette, IN 47906, USA}}
\date{(Dated: 7 July 2020)}

\begin{document}
\maketitle

\begin{abstract}
    We discuss how high energy neutrino communications could be synchronized to large-scale astrophysical events either in addition to or instead of electromagnetic signals.
\end{abstract}


\vspace{0.1in}

\begin{multicols}{2}

Understanding signal carriers is crucial to developing technology relevant to both sending and receiving messages. \textit{Project Cyclops} \cite{oliver_project_1996} defined four properties that all signal carriers should have in the context of extraterrestrial intelligence (ETI) communications: (i) a low energy per carrier; (ii) a high speed; (iii) easy generation, launching, focusing, and capturing; and (iv) minimal absorption/deflection by the interstellar medium (ISM). Since electromagnetic (EM) signals satisfy many of these properties, they have been widely explored (e.g., Ref.\ \cite{tarter_search_2001}). For long-baseline communications, EM signals in the radio wave regime of 1-10 GHz are preferable \cite{cocconi_searching_1959}. 

    
    
    

While EM signals in this frequency range seem ideal for photon communications, the high speed of neutrinos, and their weak interference with the ISM, make them potentially suitable for long-range communications (see summary of relevant literature in Ref.\ \cite{hippke_benchmarking_2018}). However, there are still other limitations. Neutrino signals should not resemble, for example, solar neutrinos, supernova neutrinos, atmospheric neutrinos, or artificial neutrinos from reactor sources. This excludes most neutrino energies below the GeV level, and hence suggests that energies should approach 1 TeV and higher.

In the search for ETI, signals transmitted over astrophysical distances should coincide with events at the Schelling point in game theory \cite{schelling_strategy_1960}. This allows the sender (here, the ETI) and the receiver (the Earth) to establish communications without prior knowledge of each other's presence. One example of the Schelling point in action is illustrated in the following scenario: Two people living in Chicago are told to meet each other in the city the following day without knowing when or where. While there are seemingly infinite times and places for the two to meet, both citizens are most likely motivated to meet at Millennium Park at noon, since both citizens could assume that anybody else in the city would view this as an important place and time. This idea is readily applied to ETI communications.

Several potential Schelling space-time points for ETI have been proposed, and one group of candidates are binary neutron star (BNS) inspirals. Since ETI might not desire to communicate within their own galaxy for their civilization's security \cite{kipping_cloaking_2016}, BNS inspirals outside a sender's galaxy would be appropriate candidates. As discussed in Ref.\ \cite{nishino_search_2018}, an extragalactic ETI civilization could transmit EM signals that are synchronized to pass a BNS system as it initiates an inspiral. As the inspiral signal travels toward a potential receiver, the EM signal would coincide with the inspiral signal. In principle, the receiver could then disentangle the ETI signal from that of the inspiral.

We wish, instead, to explore the idea of synchronizing neutrino communications to BNS inspirals in place of (or in addition to) EM signals. This is motivated by two observations of an unexpected signal \cite{fischbach_indications_2018} accompanying the GW170817 neutron star inspiral \cite{ligo_scientific_collaboration_and_virgo_collaboration_gw170817_2017}. The authors of Ref.\ \cite{fischbach_indications_2018} analyzed the unexpected signal for particles of mass $m_X$ and energy $\sim10$ MeV. They inferred a bound on $m_X$ that included an estimated neutrino mass in the eV or sub-eV range. It was subsequently shown that a second independent experiment detected the same signal \cite{fischbach_comment_2020}. There is no currently understood process that would create such a precursor signal, especially with neutrinos. In addition, there is no conventional theoretical mechanism that would predict a perturbed decay rate similar to what was reported in Refs.\ \cite{fischbach_indications_2018} and \cite{fischbach_comment_2020}.

We do note, however, that there are several reports of decay rate fluctuations correlated with changing neutrino fluxes (see Ref.\ \cite{stancil_search_2017} for summary of relevant literature). While there are questions surrounding whether these perturbations are, in fact, correlated with changing neutrino fluxes, it has been noted that these perturbations cannot be entirely explained by external effects such as temperature \cite{jenkins_analysis_2010}. 

An understanding of radioactivity would be expected of receivers \cite{whitmire_nuclear_1980}. This would be beneficial for the one-way communication intended by ETI. The multitude of ``tabletop" experiments (e.g., those similar to the apparatus in Ref.\ \cite{fischbach_indications_2018}) would likely occur simultaneously for potential receivers. The simplicity of radioactivity experiments, compared to the radio telescopes needed for EM signal receiver technology, is also advantageous. Hence, potential receivers would be expected to have the technology capable of detecting the ETI signals without committing to their search. This contrasts with what would likely be needed with EM signals, e.g., with SETI on Earth. ETI signals in decay data could resemble the unexpected signal reported by Refs.\ \cite{fischbach_indications_2018} and \cite{fischbach_comment_2020}. It is interesting to note that the inspiral signal began approximately 2 hours before the GW170817 gravity wave detection. Although the authors of Ref.\ \cite{fischbach_indications_2018} made no effort to interpret the signal, it would be natural for ETI to consider sending a precursor signal, precisely because it would be unexpected and, therefore, somewhat easier to detect.

However, the power required to transmit a signal of neutrinos that would overcome the local solar neutrino flux background at energies $\sim1$ MeV is large. It can be estimated to be on the order of the solar luminosity even for distances smaller than $1$ pc. There is additionally the concern that low-energy neutrinos released from a beam source would not be as focused as high-energy neutrinos, leading to a further energy-inefficient communication mechanism. Unless ETI had outposts nearby, such communication might be far less efficient than EM communications. On the other hand, there could be more ETI nearby than previously expected \cite{westby_astrobiological_2020}. There is also the possibility of new, yet-to-be-discovered neutrino interactions that could be employed by an advanced ETI civilization. 

The power $P_\gamma$ required for the scheme in Ref.\ \cite{nishino_search_2018} was

\begin{equation}
    P_\gamma = 0.9 \text{ TW} \left(\frac{D}{40 \text{ Mpc}}\right)^2,
\end{equation}

\noindent where $D$ is the distance between the sender and the receiver. This motivates us to consider neutrino communication schemes requiring a significantly smaller incident flux that would be readily predicted with more standard neutrino interactions. Since the neutrino interaction cross-section increases with energy---and the backgrounds at high energies fall off quickly---we might explore energies such as at the Glashow resonance at $6.3$ PeV as in Ref.\ \cite{learned_galactic_2009}, which would generally not require more than a Type I or Type II ETI civilization on the Kardashev scale \cite{kardashev_transmission_1964}.

If ETI had attempted to communicate with high-energy neutrino signals in coincidence with GW170817, we would expect to see these neutrinos with current experiments that are sensitive to these energies. In fact, there have been searches for neutrinos following GW170817 \cite{abe_search_2018, albert_search_2017}. Super-Kamiokande did not find any significant signal in the $3.5$ MeV to $\sim100$ PeV range in a $\pm500$-s window \cite{abe_search_2018}. Additionally, no neutrinos in the GeV to EeV range were found for the $\pm500$-s window coming from \textsc{Antares}, IceCube, or the Pierre Auger Observatory \cite{albert_search_2017}.

Assuming ETI exist and would choose to communicate with high-energy neutrino signals, there are two possible interpretations of these results. The first is that the ETI did not synchronize any communications with this specific event. The second is that the signal fell outside of the $\pm500$-s interval explored in both Refs.\ \cite{abe_search_2018} and \cite{albert_search_2017}. It is worth noting that ETI would be limited by uncertainties in their estimation of the exact moment of the occurrence of a BNS inspiral. As pointed out in Ref.\ \cite{nishino_search_2018}, ETI could mismatch the signal by much more than $\pm500$ s, and by as much as hours, weeks, or even months depending on their instrumental capabilities.

Regardless of whether there was an ETI signal associated with GW170817, we note that the receiver technology necessary for sensitive EM communications in the BNS inspiral scheme might not yet be available on Earth. The complexity of these large antennae may make neutrino detectors more attractive targets for ETI. As previously mentioned, ETI might not want to require a civilization to commit entirely to ETI searches. Furthermore, ETI would consider the short lifetimes for advanced civilizations \cite{ulmschneider_intelligent_2005}, which would otherwise allow for more complex SETI technology to be developed.

In summary, although radio wave transmissions are energy efficient in comparison to neutrino communications over large distances, neutrino communications are less likely to be perturbed en route. They would require receiver technology currently available to civilizations engaging in neutrino physics and astronomy, compared to the traditional large-scale commitment to ETI searches. Finally, synchronized neutrino signals with large-scale events might be a desirable target for future SETI projects.

\end{multicols}
\newpage

\begin{thebibliography}{}
\bibitem{oliver_project_1996}
Oliver, BM, and J Billingham. 1996. “Project Cyclops: A Design Study of a System for Detecting Extraterrestrial Intelligent Life. NASA Rep. CR 114445.” Little Ferry, NJ: SETI Institute.

\bibitem{tarter_search_2001}
Tarter, Jill. 2001. “The Search for Extraterrestrial Intelligence (SETI).” Annual Review of Astronomy and Astrophysics 39 (1): 511–48. https://doi.org/10.1146/annurev.astro.39.1.511.

\bibitem{cocconi_searching_1959}
Cocconi, Giuseppe, and Philip Morrison. 1959. “Searching for Interstellar Communications.” Nature 184 (4690): 844–46. https://doi.org/10.1038/184844a0.

\bibitem{hippke_benchmarking_2018}
Hippke, Michael. 2018. “Benchmarking Information Carriers.” Acta Astronautica 151 (October): 53–62. https://doi.org/10.1016/j.actaastro.2018.05.038.

\bibitem{schelling_strategy_1960}
Schelling, T. 1960. The Strategy of Conflict. Cambridge: Harvard University Press.

\bibitem{kipping_cloaking_2016}
Kipping, David M., and Alex Teachey. 2016. “A Cloaking Device for Transiting Planets.” Monthly Notices of the Royal Astronomical Society 459 (2): 1233–41. https://doi.org/10.1093/mnras/stw672.


\bibitem{nishino_search_2018}
Nishino, Yuki, and Naoki Seto. 2018. “The Search for Extra-Galactic Intelligence Signals Synchronized with Binary Neutron Star Mergers.” The Astrophysical Journal 862 (2): L21. https://doi.org/10.3847/2041-8213/aad33d.

\bibitem{fischbach_indications_2018}
Fischbach, E., V. E. Barnes, N. Cinko, J. Heim, H. B. Kaplan, D. E. Krause, J. R. Leeman, et al. 2018. “Indications of an Unexpected Signal Associated with the GW170817 Binary Neutron Star Inspiral.” Astroparticle Physics 103 (December): 1–6. https://doi.org/10.1016/j.astropartphys.2018.06.001.

\bibitem{ligo_scientific_collaboration_and_virgo_collaboration_gw170817_2017}
LIGO Scientific Collaboration and Virgo Collaboration, B. P. Abbott, R. Abbott, T.D. Abbott, F. Acernese, K. Ackley, C. Adams, et al. 2017. ``GW170817: Observation of Gravitational Waves from a Binary Neutron Star Inspiral." Physical Review Letters 119 (16): 161101. https://doi.org/10.1103/PhysRevLett.119.161101.

\bibitem{fischbach_comment_2020}
Fischbach, E., D. E. Krause, and M. Pattermann. 2020. “Comment on 'Testing Claims of the GW170817 Binary Neutron Star Inspiral Affecting $\beta$-Decay Rates’.” ArXiv:2003.00092 [Nucl-Ex], February. http://arxiv.org/abs/2003.00092.

\bibitem{stancil_search_2017}
Stancil, Daniel D., Sümeyra Balci Yegen, David A. Dickey, and Chris R. Gould. 2017. “Search for Possible Solar Influences in Ra-226 Decays.” Results in Physics 7 (January): 385–406. https://doi.org/10.1016/j.rinp.2016.12.051.

\bibitem{jenkins_analysis_2010}
Jenkins, Jere H., Daniel W. Mundy, and Ephraim Fischbach. 2010. “Analysis of Environmental Influences in Nuclear Half-Life Measurements Exhibiting Time-Dependent Decay Rates.” Nuclear Instruments and Methods in Physics Research Section A: Accelerators, Spectrometers, Detectors and Associated Equipment 620 (2): 332–42. https://doi.org/10.1016/j.nima.2010.03.129.

\bibitem{whitmire_nuclear_1980}
Whitmire, Daniel P., and David P. Wright. 1980. “Nuclear Waste Spectrum as Evidence of Technological Extraterrestrial Civilizations.” Icarus 42 (1): 149–56. https://doi.org/10.1016/0019-1035(80)90253-5.

\bibitem{westby_astrobiological_2020}
Westby, Tom, and Christopher J. Conselice. 2020. “The Astrobiological Copernican Weak and Strong Limits for Extraterrestrial Intelligent Life.” The Astrophysical Journal 896 (1): 58. https://doi.org/10.3847/1538-4357/ab8225.

\bibitem{learned_galactic_2009}
Learned, John G., Sandip Pakvasa, and A. Zee. 2009. “Galactic Neutrino Communication.” Physics Letters B 671 (1): 15–19. https://doi.org/10.1016/j.physletb.2008.11.057.

\bibitem{kardashev_transmission_1964}
Kardashev, N. S. 1964. “Transmission of Information by Extraterrestrial Civilizations.” Soviet Astronomy 9: 217.

\bibitem{abe_search_2018}
Abe, K., C. Bronner, Y. Hayato, M. Ikeda, K. Iyogi, J. Kameda, Y. Kato, et al. 2018. “Search for Neutrinos in Super-Kamiokande Associated with the GW170817 Neutron-Star Merger.” The Astrophysical Journal 857 (1): L4. https://doi.org/10.3847/2041-8213/aabaca.

\bibitem{albert_search_2017}
Albert, A., M. Andre, M. Anghinolfi, M. Ardid, J.-J. Aubert, J. Aublin, T. Avgitas, et al. 2017. “Search for High-Energy Neutrinos from Binary Neutron Star Merger GW170817 with ANTARES, IceCube, and the Pierre Auger Observatory.” The Astrophysical Journal 850 (2): L35. https://doi.org/10.3847/2041-8213/aa9aed.

\bibitem{ulmschneider_intelligent_2005}
Ulmschneider, Peter. 2005. Intelligent Life in the Universe: Principles and Requirements Behind Its Emergence. Springer Science \& Business Media.


\end{thebibliography}

\end{document}